\documentclass[a4paper,11pt]{article}
\usepackage{pos}
\usepackage{psfrag}
\usepackage{multirow}

\title{Exclusive photoproduction of a photon-meson pair: A new class of observables to probe GPDs}

\author[1]{Goran Duplan\v{c}i\'{c}}
\author*[2]{Saad Nabeebaccus}
\author[1]{Kornelija Passek-K}
\author[3]{Bernard Pire}
\author[4,5]{Jakob Sch\"onleber}
\author[6]{Lech Szymanowski}
\author[2]{Samuel Wallon}

\affiliation[1]{Theoretical Physics Division, Rudjer Bo{\v s}kovi{\'c} Institute,
	HR-10002 Zagreb, Croatia}
\affiliation[2]{Universit\'e Paris-Saclay, CNRS/IN2P3, IJCLab, 91405 Orsay, France}
\affiliation[3]{CPHT, CNRS, Ecole polytechnique, Institut Polytechnique de Paris, 91128 Palaiseau, France}
	\affiliation[4]{Institut f\"ur Theoretische Physik, Universit\"at Regensburg, D-93040 Regensburg, Germany}
\affiliation[5]{RIKEN BNL Research Center, Brookhaven National Laboratory, Upton, NY 11973, USA}
\affiliation[6]{National Centre for Nuclear Research (NCBJ), 02-093 Warsaw, Poland}

\abstract{We discuss the exclusive photoproduction of a photon-meson pair with large $ p_{T} $ as a channel to probe generalised parton distributions (GPDs). Like other $ 2\to3 $ exclusive processes, this channel allows us to better study the $ x $-dependence of GPDs, in contrast to $ 2\to2 $ processes such as Deeply Virtual Compton Scattering (DVCS) which give only ``moment-type'' information. Moreover, it also gives the possibility to access, at leading twist, the chiral-odd GPDs, which are still completely unknown experimentally. In the first part of these proceedings, we present the computation of the amplitude at leading order and leading twist, for charged pions, and rho mesons of any charge and polarisation. These processes are sensitive to quark GPDs only. We also discuss the possibility of measuring such processes at various experiments, namely JLab, COMPASS, future EIC and LHC in ultra-peripheral collisions. In particular, in collider experiments, the dependence of GPDs at small skewness $ (<10^{-3}) $ can be extracted. For the second part, we report on our recent work on the exclusive photoproduction of $ \pi^{0}\gamma  $ pair, which, in addition to a quark GPD channel, also has a contribution from gluon GPDs. In the latter case, we demonstrate that a collinear factorisation of the process fails, due to the presence of a Glauber pinch. Our findings also imply that other similar processes, like $  \pi ^{0} N \to \gamma \gamma N $ also suffer from the same issue. On the other hand, we stress that the processes discussed earlier, which are sensitive to the quark GPD channels only, are safe from these factorisation breaking effects.}

\FullConference{25th International Spin Physics Symposium (SPIN 2023)\\
 24-29 September 2023\\
 Durham, NC, USA\\}

\begin{document}
\maketitle

\section{Introduction}

Quark (gluon) GPDs are obtained from the non-forward matrix elements of quark-antiquark operators (gluon field strength operators) separated by a lightcone distance. At the leading twist-2, there are a total of 8 quark GPDs: 4 chiral-even, and 4 chiral-odd (helicity flip). The chiral-even ones are denoted by $ H_{q},\, E_{q},\,\tilde{H}_{q}  $ and $  \tilde{E}_{q}  $, and the chiral-odd (helicity-flip) ones by $ H_{q}^{T},\, \tilde{H}^{T}_{q},\,E_{q}^{T}  $ and $  \tilde{E}_{q}^{T}  $. Their definitions, including the ones for gluon GPDs, can be found in the review \cite{Diehl:2003ny}. A GPD is a function of 3 variables: $ x $, which corresponds to the average fraction of the nucleon longitudinal momentum that is carried by the quark/antiquark probed, $  \xi  $, the so-called skewness parameter, which corresponds to the difference in longitudinal momentum between the quark/antiquark probed (or between the two gluons probed in the case of gluon GPDs), and $ t $ which is the classic Mandelstam variable corresponding to the momentum difference squared between the incoming and outgoing nucleon.
GPDs shed light on hadron structure, since their Fourier transform (wrt $ t $) when $  \xi =0 $ gives the probability distribution of partons in the transverse plane \cite{Burkardt:2000za,Burkardt:2002hr}.

A well-studied process that is sensitive to GPDs is \textit{deeply-virtual meson production} (DVMP) which essentially corresponds to DVCS with the outgoing photon replaced by a meson \cite{Collins:1996fb,Radyushkin:1996ru}. In this case, the amplitude factorises, for specific polarisations of the incoming photon and outgoing meson, in terms of a coefficient function, a GPD, and a distribution amplitude (DA) for the outgoing meson.

In order to probe chiral-odd GPDs, since QED and QCD are chiral-even theories, one needs to consider a process where there is an even number of chiral-odd structures - otherwise, the amplitude vanishes identically when performing traces of Dirac matrices. The exclusive electroproduction of a transversely polarised $  \rho_{T}  $-meson seems to be a good candidate for this, since the DA of the $  \rho  $-meson at the leading twist-2 is chiral-odd. However, it has been shown that this amplitude vanishes identically at all orders in $  \alpha _{s} $ for such a process, since it would require a helicity transfer of 2 units from a photon \cite{Diehl:1998pd,Collins:1999un}. This vanishing only happens with a twist-2 DA. When using a twist-3 DA however, one faces issues with \textit{end-point singularities}.

A way to probe chiral-odd GPDs at the leading twist was proposed in \cite{
	ElBeiyad:2010pji} by considering the photo- and electro-production of two mesons, through a 3-body final state. The photoproduction of a $ \gamma  \rho _{T} $ pair was proposed in \cite{Boussarie:2016qop} as yet another channel, with the large (and almost opposite) transverse momenta of the photon and meson in the final state providing the hard scale for collinear factorisation of the process \cite{Qiu:2022bpq,Qiu:2022pla}. This channel led to the culmination of a series of works \cite{Duplancic:2018bum,Duplancic:2022ffo,Duplancic:2023kwe} on the photoproduction of a photon-meson pair, with the meson in the final state being either a charged pion, or a $  \rho  $-meson of any charge and polarisation. Section \ref{sec:photon-meson-photoproduction} of these proceedings are based on these works, which we point out are sensitive to quark GPDs only, since the gluon GPD channel is forbidden either on the basis of electric charge conservation, or by charge conjugation symmetry in the case of $  \rho ^{0} $.

A significant advantage in considering exclusive processes involving 3-body final states, like the ones mentioned above, is that they give access to an enhanced $ x $-dependence of GPDs, compared to exclusive processes involving 2-body final states, such as DVCS, which only effectively probe the GPD at $ x = \pm \xi $ \cite{Qiu:2023mrm}.

While a proof of factorisation has been made available for such a class of $ 2\to 3 $ exclusive processes in \cite{Qiu:2022bpq,Qiu:2022pla}, a direct calculation of the gluon GPD contribution to the exclusive photoproduction of $  \pi ^{0}\gamma  $ pair shows that the amplitude diverges at leading order and leading twist. In \cite{Nabeebaccus:2023rzr}, we explain that this divergence is a consequence of the breakdown of collinear factorisation, for such processes where the 2-gluon exchange is allowed. This is caused by the existence of a \textit{Glauber pinch} in this case. Section \ref{sec:breakdown-collinear-factorisation} of these proceedings involves a discussion of this work.

\section{Photoproduction of $  \gamma \pi ^{\pm} $ and $ \gamma  \rho _{L,T}^{0,\,\pm} $ with large invariant mass}

\label{sec:photon-meson-photoproduction}

\subsection{Kinematics}

Let us denote our process through:
\begin{align}
	\label{eq:process}
	\gamma (q) + N(p_N) \to \gamma (k) + M (p_{M}) + N' (p_{N'})\,.
\end{align}

The momenta of the particles are parametrised as follows. First, we decompose all momenta in a \textit{Sudakov basis}, such that a generic momentum $ r $ can be written as
\begin{align}
	r^{ \mu } = a p^{  \mu } + b n^{ \mu } + r^{ \mu }_{\perp}\,,
\end{align}
where the lightcone directions $ p^{ \mu } $ and $ n^{ \mu } $ are given by
\begin{align}
	p^{ \mu } = \frac{\sqrt{s}}{2} \left( 1,0,0,1 \right) \,,\qquad n^{ \mu } = \frac{\sqrt{s}}{2} \left( 1,0,0,-1 \right) \,,
\end{align}
with $ n  \cdot p = \frac{s}{2} $. We use the convention $ r_{\perp}^2 \equiv - |\vec{r}_{t}|^2 $. The momenta of the particles in process \eqref{eq:process} are then given by (with $  \Delta  = p_{N'}-p_{N} $),
\begin{align}
p_{N} &=  \left( 1+ \xi \right) p^{ \mu } + \frac{m_{N}^2}{s \left( 1+\xi \right) }n^{ \mu }\,,\quad\;\;
p_{N'}=\left( 1- \xi \right) p^{ \mu } + \frac{m_{N}^2+ |\vec{\Delta} _{t}|^2}{s \left( 1-\xi \right) }n^{ \mu }+ \Delta _{\perp}^{ \mu }\,,\quad\;\;
q^{ \mu }  = n^{ \mu }\,,\\[5pt]
\label{eq:kinematics-with-pperp}
p_{ M }^{ \mu } &=  \alpha_{ M } n^{ \mu } + \frac{  |\vec{p}_{t}+ \frac{\vec{\Delta} _{t}}{2}|^2  +m_{ M }^{2}}{ \alpha _{M}s}p^{ \mu }-p_{\perp}^{ \mu }-\frac{ \Delta _{\perp}^{ \mu }}{2}\,,\quad\;\;
k^{ \mu } =  \alpha n^{ \mu } + \frac{  |\vec{p}_{t}- \frac{\vec{\Delta} _{t}}{2}|^2  }{ \alpha s}p^{ \mu }+p_{\perp}^{ \mu }-\frac{ \Delta _{\perp}^{ \mu }}{2}\,.
\end{align}
In the above, $ m_{N} $ is the nucleon mass, $ m_{M} $ is the meson mass, and $ \xi $ is the skewness parameter.

The centre-of-mass energy of the system, $ S_{\gamma N} $, is given by
\begin{align}
	S_{\gamma N} =  \left( q+p_{N} \right) ^2 =  \left( 1+\xi \right) s + m_{N}^2\,.
\end{align}
Furthermore, we define the following useful Mandelstam variables:
\begin{align}
	t&=  \left( p_{N}-p_{N'} \right) ^2\,,\;\;\;\;u'=  \left( p_{M}-q \right) ^2\,,\;\;\;\;	t'= \left( k-q \right) ^2\,,\;\;\;\;	M_{\gamma M}^2 =  \left( p_{M}+k \right)^2\,.
\end{align}
In the Bjorken limit, where $  \vec{\Delta}_{t}  $, $ m_{N} $ and $ m_{M} $ are taken to zero, the kinematics simplifies to
\begin{align}
	\label{eq:Bjorken-kinematics}
	t = 0\,,\;\;\;
	 t' =  \bar{ \alpha }M_{\gamma M}^{2}\,,\;\;\;
	  \alpha_{ M} =\bar{ \alpha } \,,\;\;\;
	    \alpha  = \frac{-u'}{M^{2}_{\gamma M}}\,,\;\;\;
	    | \vec{p}_{t}|^2 =  \alpha \bar{ \alpha } M^{2}_{\gamma M}\,,\;\;\;
	      \xi = \frac{M_{\gamma M}^{2}}{2  S_{\gamma N}-M^{2}_{\gamma M} }\,,
\end{align}
where $  \bar{\alpha} \equiv 1- \alpha   $.
In the calculation of the amplitudes and cross sections, $ u' $ and $ M_{\gamma M}^2 $ are taken as independent variables; supplemented by $ S_{\gamma N} $, this fully specified the kinematics.

For the collinear factorisation of the process, $  \vec{p}_{t}  $ needs to be large. In practice, we impose cuts on the Mandelstam variables
\begin{align}
	-u',\,-t'>1\,  \mathrm{GeV}^2\,,\qquad  -t < 0.5\,  \mathrm{GeV}^{2}\,, 
\end{align}
in order to avoid resonances between the outgoing meson and nucleon. It has been shown in \cite{Duplancic:2022ffo} that these cuts are sufficient to ensure large $  \vec{p} _{t} $.
\subsection{Modelling of GPDs and DAs}

\label{sec:GPD-DA-models}

The GPDs we use for the numerics are parametrised in terms of double distributions \cite{Radyushkin:1998es}. For the polarised (transversity) PDFs, which are used to construct the  $  \tilde{H} _{q} $ ($ H_{q}^{T} $) GPDs, two scenarios are proposed for the parameterisation:
\begin{itemize}
	\item ``standard'' scenario, with flavour-symmetric light sea quark and antiquark distributions.
	\item ``valence'' scenario, where the densities are taken to be completely flavour anti-symmetric.
\end{itemize}

For the DAs, we use two different models:
\begin{align}
	\label{eq:DA-models}
	 \phi_{ \mathrm{asy} } (z) = 6 z  \left( 1-z \right) \,,\qquad \phi_{ \mathrm{hol} } (z) = \frac{8}{ \pi } \sqrt{z  \left( 1-z \right) }\,.
\end{align}
The first, $  \phi _{ \mathrm{asy} }(z) $ is the classic \textit{asymptotic} form, while the second, $  \phi _{ \mathrm{hol} }(z) $, is the so-called `holographic' form, 
suggested by AdS-QCD correspondence \cite{Brodsky:2006uqa}, dynamical chiral symmetry breaking on the light-front \cite{Shi:2015esa}, and recent lattice results \cite{Gao:2022vyh}, where an even flatter form  was obtained.

More details regarding the parameterisation of the DA and GPD can be found in \cite{Duplancic:2022ffo,Duplancic:2023kwe}.

\subsection{Observables}

The amplitude $  {\cal A }  $ is computed through the convolution of the coefficient function $ T(x,z) $, the GPD $ H(x, \xi ,t) $, and the DA $  \phi (z) $,
\begin{align}
 {\cal A } = \int_{-1}^{1}dx \int _{0}^{1}dz\,T(x,z)\,H(x, \xi ,t)\, \phi (z)\,.
\end{align}
For the DA models mentioned in \eqref{eq:DA-models}, the integrals over the momentum fraction $ z $ are performed \textit{analytically}. The remaining integral over the momentum fraction $ x $ is then performed \textit{numerically}. 

In practice, for the chiral-even amplitude, we only keep the contributions from the GPDs $ H_{q} $ and $  \tilde{H}_{q}  $, while for the chiral-odd amplitude, we only keep those from $ H_{q}^{T} $, since the contributions from the other GPDs are kinematically suppressed. Further details regarding the computation of the amplitudes can be found in \cite{Duplancic:2022ffo} for the charged pion case, and in \cite{Duplancic:2023kwe} for the  $ \rho  $ meson case.

\subsubsection{Unpolarised cross sections}

\begin{figure}
\centering
\psfrag{HHH}{\hspace{-1.5cm}\raisebox{-.5cm}{\scalebox{.7}{$M^2_{\gamma \pi^{+}} ({\rm GeV}^{2})$}}}
\psfrag{VVV}{\raisebox{.3cm}{\scalebox{.7}{$\hspace{-.7cm}\displaystyle\frac{d \sigma^{\rm even}_{  \gamma \pi^{+} }}{d M^2_{\gamma \pi^{+}}}~({\rm pb} \cdot {\rm GeV}^{-2})$}}}
\psfrag{TTT}{}
\hspace{0.33cm}
\includegraphics[width=0.46\textwidth]{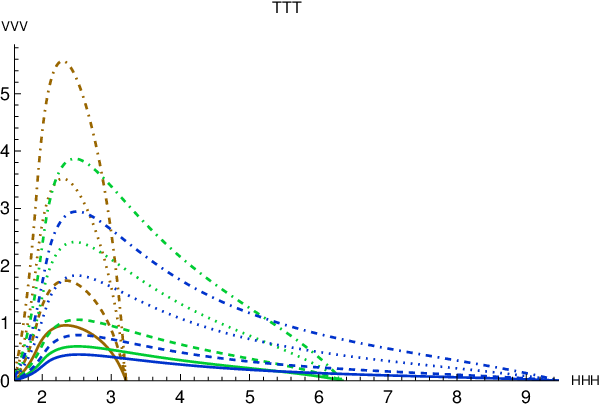}
\hfill
\psfrag{TTT}{}
\psfrag{HHH}{\hspace{-1.5cm}\raisebox{-.5cm}{\scalebox{.7}{$S_{\gamma N} ({\rm GeV}^{2})$}}}
\psfrag{VVV}{\raisebox{.3cm}{\scalebox{.7}{$\hspace{-.4cm}\displaystyle\sigma^{even}_{{\gamma \pi^{+}}}~({\rm pb})$}}}
\includegraphics[width=0.46\textwidth]{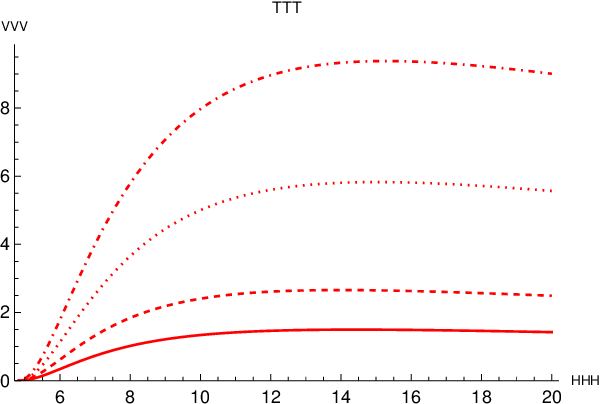}
\caption{In both plots, the outgoing meson is a $  \pi ^{+} $. Dashed (non-dashed) corresponds to holographic (asymptotical) DA, while dotted (non-dotted) corresponds to standard (valence) scenario for the GPD model used. \textbf{Left}: The differential cross section wrt $ M_{\gamma M}^{2} $ as a function of $ M_{\gamma  M}^{2} $ is shown for different $ S_{\gamma N} = 8\,  \left( \mathrm{Brown} \right)  ,\, 14\,( \mathrm{Green} ), \, 20\,( \mathrm{Blue} )\; \mathrm{GeV}^{2} $. \textbf{Right}: The cross section is plotted as a function of $ S_{\gamma N} $. }
\label{fig:cross-section}
\end{figure}

We denote the squared amplitude, after summing over the helicities of the incoming and outgoing particles, by $ |\overline{ {\cal A } }|^2 $. Then, the unpolarised differential cross section can be written as
\begin{align}
	\frac{d \sigma }{dt\, du'\, dM_{\gamma M}^{2}}\Bigg{|}_{(-t)=(-t)_{ \mathrm{min} }} = \frac{|\overline{ {\cal A } }|^2}{32 S_{\gamma N}^{2} M_{\gamma M}^{2} \left( 2 \pi  \right)^3 }\,,
\end{align}
where $ (-t)_{ \mathrm{min} }  = -\frac{4 \xi ^2 m_{N}^2}{1- \xi ^2}$. Figure \ref{fig:cross-section} shows the differential cross section wrt $ M_{\gamma M}^{2} $ as a function of $ M_{\gamma M}^{2} $ (left), and the cross section as a function of $ S_{\gamma N} $ (right), for a produced $  \pi ^{+} $ meson. For illustrative purposes, we only show the plots that correspond to JLab kinematics (i.e.~$ S_{\gamma N,\, \mathrm{max} } \approx 20\,  \mathrm{GeV}^2  $). Thus, one concludes that the GPD model used has a significant effect on the cross section (compare dotted vs non-dotted lines) for $  \pi ^{+}\gamma  $ photoproduction, which means this channel can be used to disentangle between the two GPD models considered. Moreover, one also finds that the cross section is larger with the holographic form of the DA.

We refer the reader to the original papers \cite{Duplancic:2022ffo,Duplancic:2023kwe} for plots corresponding to higher centre-of-mass energies, as well as other mesons.

\subsubsection{Polarisation asymmetries with respect to the incoming photon}

We also computed polarisation asymmetries wrt the incoming photon. By a direct calculation, we found that the \textit{circular} polarisation asymmetry vanishes for all the mesons that we considered. This vanishing is linked to the invariance of QCD/QED under parity, for more details, see \cite{Duplancic:2022ffo}. We therefore calculated the \textit{linear} polarisation asymmetry (LPA), using the \textit{Kleiss-Stirling spinor techniques}. It turns out that the linear polarisation asymmetry also vanishes for the case of a transversely polarised $  \rho  $-meson. Such a LPA is calculated by
\begin{align}
	\label{eq:LPA-def}
	 \mathrm{LPA} = \frac{d \sigma _{x}-d \sigma _{y}}{d \sigma _{x}+d \sigma _{y}}\,,
\end{align}
where $ d \sigma  $ represents a differential or integrated cross section as appropriate, and the subscript $ x $ (or $ y $) represents the direction of polarisation of the incoming photon, in a frame where the $ x $-direction is defined by $ p_{\perp} $ in \eqref{eq:kinematics-with-pperp} (i.e.~the direction of the outgoing photon defines $ x $). In practice, the measured asymmetry in the lab frame, which we denote by $  \mathrm{LPA}_{ \mathrm{Lab} }  $ is related to the above LPA in \eqref{eq:LPA-def} by
\begin{align}
\mathrm{LPA}_{ \mathrm{Lab} } = \mathrm{LPA}\cos  \left( 2 \theta  \right) \,,
\end{align}
where $  \theta  $ is the angle between the lab frame $ x $-direction and $ p_{\perp} $, which varies event by event.

\begin{figure}
	\centering
	\psfrag{HHH}{\hspace{-1.5cm}\raisebox{-.5cm}{\scalebox{.8}{$ M^{2}_{\gamma  \rho ^{+}} ({\rm 
				GeV}^{2})$}}}
\psfrag{VVV}{ $ \mathrm{LPA}_{{  \gamma { \rho }^{+}_{L} }} $ }
\psfrag{TTT}{}
	\includegraphics[width=0.46\textwidth]{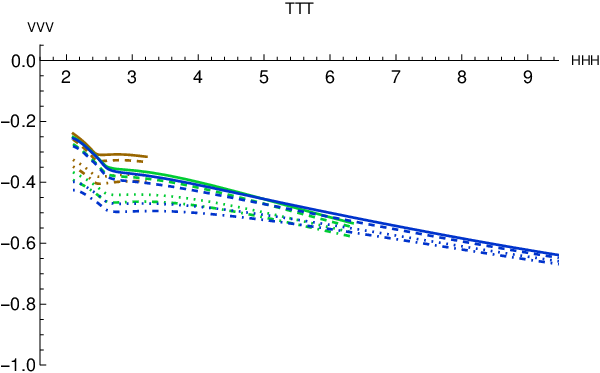}
	\hfill
	\psfrag{HHH}{\hspace{-1.5cm}\raisebox{-.5cm}{\scalebox{.8}{$M^{2}_{\gamma  \pi ^{+}} ({\rm 
					GeV}^{2})$}}}
	\psfrag{VVV}{$ \mathrm{LPA}_{{ \gamma { \pi ^{+}} }} $}
	\psfrag{TTT}{}
	\includegraphics[width=0.46\textwidth]{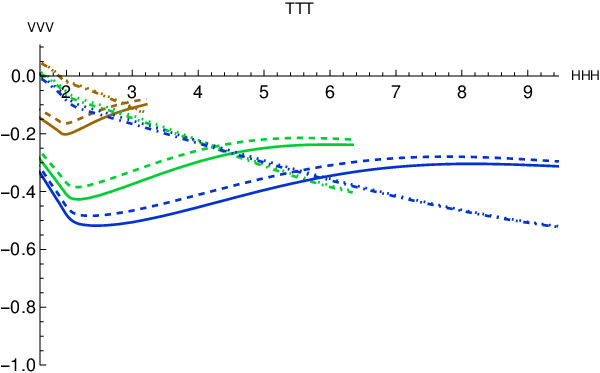}
	\caption{ The LPA, constructed from differential cross sections wrt $ M_{\gamma M}^{2} $, is shown as a function of $ M_{\gamma M}^{2} $ for the $   \rho ^{+}  $ (left) and $  \pi ^{+} $ cases. The meanings of the different colours and types of lines are exactly the same as those in Figure \ref{fig:cross-section}.}
	\label{fig:LPA}
\end{figure}

Figure \ref{fig:LPA} shows the variation of LPA, constructed from the differential cross section in $ M_{\gamma M}^{2} $, as a function of $ M_{\gamma M}^2 $ for the $  \rho ^{+}_{L} $ meson (left) and $  \pi ^{+} $ meson (right) cases. Interestingly, the plots show that, depending on the meson produced, the two GPD models either have no effect on the LPA (for the $  \rho ^{+} $ case), or change the variation of the LPA completely (for the $  \pi ^{+} $ case). This suggests the use of the LPA for disentangling GPD models, where in particular, the effect coming from the DA models is suppressed. Moreover, the LPA is sizeable in both cases, reaching up to $ 60\% $.

\subsection{Expected number of events at various experiments}

We finally performed a full phase space integration, taking into account the photon flux, so as to obtain the estimated number of events for the different mesons we consider, at various experiments. In particular, we calculate the numbers for JLab (taking the integrated luminosity $ \int  {\cal L }dt  = 864 \, \mathrm{fb}^{-1}   $), future EIC (taking  $ \int  {\cal L }dt  = 10\; \mathrm{fb}^{-1}  $) and LHC in ultraperipheral collisions (UPCs) (assuming $ \int  {\cal L }dt  = 1200   \mathrm{nb}^{-1}  $). The numbers are shown in Table \ref{tab:number-events}, for a proton target. One can exploit the high energies available in a collider environment in order to perform a \textit{small $  \xi  $} study of quark GPDs, by restricting $ 300 < \frac{S_{\gamma N}}{\mathrm{GeV}^{2}} < 20000 $, which roughly\footnote{The correspondence between $ S_{\gamma N}$ and $  \xi   $ is given in \eqref{eq:Bjorken-kinematics}, which involves $ M_{\gamma M}^{2} $. In estimating the values of  $ \xi  $, we used the value of $ M_{\gamma M}^{2} $ at the peak in the cross section plots in Figure \ref{fig:cross-section}, which remains at the same position of  $ \sim $ 2-3  GeV$ ^{2} $. } translates to $ 5 \cdot 10^{-5} <  \xi  < 5 \cdot 10^{-3} $. The expected number of events by employing this cut is also found in Table \ref{tab:number-events}.

\begin{table}
	\centering
\begin{tabular}{|c|c|c|c|}
\hline
Experiment & Meson & Number of events & Number of events with $ S_{\gamma N}>300  \mathrm{GeV}^2  $ \\
\hline
\hline
\multirow{3}{*}{JLab}
& $  \rho ^{0}_{L} $ & 1.3-2.4 $ \times 10^{5}$ & - \\\cline{2-4}
&$  \rho ^{0}_{T} $ & 2.1-4.2 $ \times 10^{4} $ & - \\\cline{2-4}
&$  \pi ^{+} $ &  0.3-1.8 $\times 10^{5}$ & - \\\hline
\multirow{3}{*}{EIC}
& $  \rho ^{0}_{L} $ & 1.3-2.4 $ \times 10^{4} $ & 0.6-1.2 $ \times 10^{3} $ \\\cline{2-4}
&$  \rho ^{0}_{T} $ & 1.2-2.4 $ \times 10^{3} $ & - \\\cline{2-4}
&$  \pi ^{+} $ & 0.2-1.3 $ \times 10^{4} $ & 1.4-5.0 $ \times 10^{2} $ \\\hline
\multirow{3}{*}{LHC in UPCs}
& $  \rho ^{0}_{L} $ & 0.9-1.6 $ \times 10^{4} $ & 4.1-8.1 $ \times 10^{2} $ \\\cline{2-4}
&$  \rho ^{0}_{T} $ & 0.8-1.7 $ \times 10^{3} $ & - \\\cline{2-4}
&$  \pi ^{+} $ & 1.6-9.3 $ \times 10^{3} $ & 1.0-3.4 $ \times 10^{2} $ \\\hline
\end{tabular}
\caption{The expected number of events for the photoproduction of a photon-meson at JLab, future EIC and LHC in UPCs is shown for the $  \rho ^{0}_{L} $, $  \rho ^{0}_{T} $ and $  \pi ^{+} $ cases, on a proton target. The number of events at small skewness $  \xi  $ ($ S_{\gamma N}> 300  \mathrm{GeV}^{2}  $) is also shown for EIC and LHC kinematics.}
\label{tab:number-events}
\end{table}

Thus, one finds that the statistics are very good, which warrants a proper experimental analysis of the process, especially at JLab, where the number of events can be as high as $ 10^5 $. Furthermore, we performed an analysis based on the detection of the outgoing photon at JLab, and found that the numbers remain more or less the same when such a cut is applied \cite{Duplancic:2023kwe}. Finally, we note that for the chiral-odd case ($  \rho _{T} $), the cross section is proportional to $  \xi ^2 $, which implies that the expected number of events becomes negligible at small $  \xi  $, and hence they are omitted from the table.

\section{Breakdown of collinear factorisation in exclusive $  \pi ^{0}\gamma  $ pair photoproduction}

\label{sec:breakdown-collinear-factorisation}

In this section, we discuss our recent work in \cite{Nabeebaccus:2023rzr} regarding collinear factorisation breaking effects in the exclusive photoproduction of a $  \pi ^{0}\gamma  $ pair. In contrast to the previously discussed mesons in the final state, namely charged pions and $  \rho  $-mesons of any charge and polarisation, this process is sensitive to the exchange of 2 gluons in the $ t $-channel. 

\begin{figure}
\centering
\includegraphics[width=0.4\textwidth]{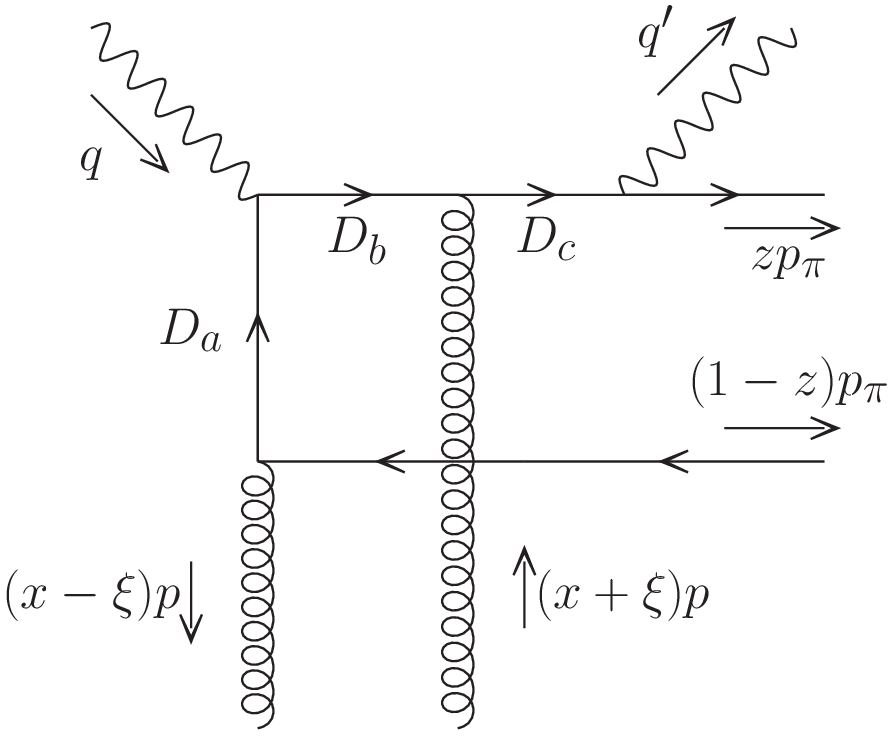}
\hfill
\includegraphics[width=0.55\textwidth]{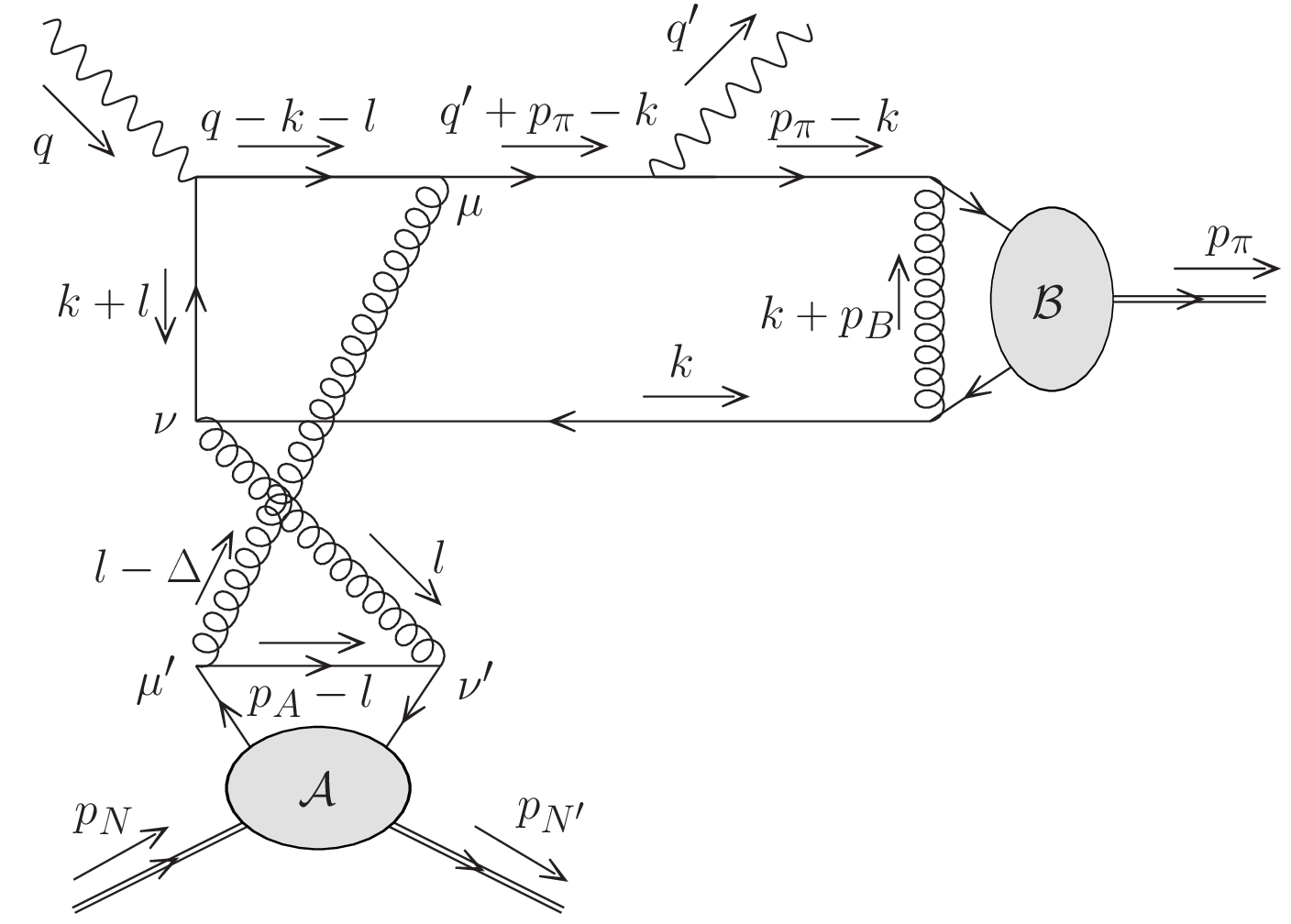}
\caption{\textbf{Left}: A diagram that diverges upon convolution with the gluon GPD and the pion DA. The indices $ i $ and $ j $ represent the transverse Lorentz indices of the two gluons, and the propagators are denoted by $ D $ with a subscript. \textbf{Right}: A two-loop diagram corresponding to our process, whose different regions of loop integration can be studied.}
\label{fig:problematic-diagram}
\end{figure}

Assuming collinear factorisation, a direct calculation, at leading order and leading twist, leads to a \textit{divergent} amplitude. The diagrams that are responsible for this divergence are those of the type shown in left side of Figure \ref{fig:problematic-diagram}. These correspond to cases where the \textit{incoming} photon attaches between the two gluons that go to the GPD (independent of the way the outgoing photon attaches to the quark line). In fact, the divergence can be shown to be \textit{purely imaginary}, and comes from the pinching of two propagator poles.\footnote{We point out that such a topology in the quark GPD case does not exist, since for this to happen, the incoming photon would have to couple to the gluon that connects the two quark lines that go into the quark GPD, which is impossible.} In the specific case\footnote{Note that the outgoing photon momentum is now $ q' $, since we will later use $ k $ for the loop momentum of the quark.} shown in the left of Figure \ref{fig:problematic-diagram}, these correspond to $ D_{a} $ and $ D_{b} $. The relevant integration region that causes the divergence is $ x \to  \xi  $ and $ \bar{z} \to 0 $, and in this limit, the amplitude can be shown to be proportional to
\begin{align}
	\label{eq:problematic-diagram-integral}
	\int_{-1}^{1} dx \int _{0}^{1}dz\frac{1}{ \left[  \left( x-\xi \right)+A \bar{z} - i\epsilon   \right]\left[ x-\xi + i\epsilon  \right]  }\,,
\end{align}
where $ A $ is a positive number fixed by the external kinematics. The denominator $ \left[ x-\xi + i\epsilon  \right]  $ corresponds to $ D_{a} $, while $ \left[  \left( x-\xi \right)+A \bar{z} - i\epsilon   \right] $ corresponds to $ D_{b} $. The two propagators have opposite $ i\epsilon  $ prescriptions, causing the pinching and the divergence of the integral in \eqref{eq:problematic-diagram-integral}.

After a careful analysis of the process, we discovered that this divergence is in fact a consequence of the breakdown of collinear factorisation. This is caused by the existence of a \textit{Glauber pinch}, which contributes to the same power in $ Q $ in the twist expansion of the amplitude as the usual \textit{collinear pinch}. Consider the two-loop diagram shown in the right of Figure \ref{fig:problematic-diagram}, where the momentum $  \Delta  = p_{N'}-p_{N} $, and $ k $ and $ l $ are loop momenta. First, let us introduce the idea of \textit{momentum scaling}. We denote by $ Q $ a hard scale in the process (for example, the large transverse momentum $ | \vec{p} _{t}| $ of the outgoing photon). Then, defining the small parameter $  \lambda  $, such that a generic soft scale is obtained through $  \lambda Q \sim  \Lambda _{ \mathrm{QCD} },\,m_{N},\,m_{\pi},\,\sqrt{|t|} $, we have, for a generic momentum $ r $ in the Sudakov basis,
\begin{align}
&r \sim Q  \left(1, \lambda ^2, \lambda   \right) &&\textrm{(coll. to $ + $ direction)}\,,
&&r \sim Q  \left( \lambda ^2, 1,\lambda   \right) &&\textrm{(coll. to $ - $ direction)}\,,\\
&r \sim Q  \left( \lambda ^2,  \lambda ^{2},\lambda^{2}   \right) &&\textrm{(ultrasoft)}\,,
&&r \sim Q  \left( \lambda ,  \lambda ,\lambda   \right) &&\textrm{(soft)}\,.
\end{align}
The usual convention is such that \textit{soft} and \textit{ultrasoft scalings} correspond to cases where all momentum components scale in exactly the same way. This basically means that the particle is \textit{on-shell}. There is, however, a special type of soft scaling, called \textit{Glauber scaling}, such that the transverse momentum component is much larger than the $ + $ and $ - $ components of momentum,
 \begin{align}
	r^{+}r^{-} \ll |r_{\perp}|^2 \implies r \sim Q  \left(  \lambda ^2, \lambda ^2, \lambda  \right) ,\,r \sim Q  \left(  \lambda , \lambda ^2, \lambda  \right) ,\,\textrm{etc.},
\end{align}  
 such that the particle is now \textit{off-shell}.

In the analysis of the two-loop diagram on the right of Figure \ref{fig:problematic-diagram}, we fix the external kinematics such that $ p_{N},\,p_{N'},\, \Delta ,\,p_{A} $ are collinear to the $ + $ direction, $ p_{\pi},\,p_{B}$ are collinear to the $ - $ direction, and $ q,\,q' $ correspond to collinear particles in an arbitrary direction (i.e.~their scalings are $ Q \left( 1,1,1 \right)  $, with the condition that $ q^2 = q'^2 =0 $). The blob $  {\cal A }  $ ($  {\cal B }  $) corresponds to sub-diagrams where all momenta are collinear to the nucleon (pion) sector. One can now carefully study the different regions of loop integration for $ k $ and $ l $. The classic collinear pinch is given by the configurations
\begin{align}
	\label{eq:collinear-pinch}
	l \sim Q  \left( 1, \lambda ^{2} , \lambda \right) \,,\qquad k \sim Q  \left(  \lambda ^2,1, \lambda  \right) \,,
\end{align}
i.e.~when the gluon loop momentum $ l $ is collinear to the nucleon momenta and the quark loop momentum $ k $ is collinear to the pion momentum. In this case, the $ l^{-} $ component of the gluon loop momentum, and the $ k^{+} $ component of the quark loop momentum are pinched. Moreover, a power counting analysis of the collinear pinch shows that the amplitude has a scaling of $  \lambda ^{1} $ \cite{Nabeebaccus:2023rzr}.

We now demonstrate the Glauber pinch that we discovered explicitly. We consider the scalings of the loop momenta to be
\begin{align}
l \sim Q  \left(  \lambda , \lambda ^2, \lambda  \right) ,\,\qquad k \sim Q  \left(  \lambda , \lambda , \lambda  \right) ,\,
\end{align}
i.e.~a Glauber scaling for $ l $ and a soft scaling for $ k $. Consider now the two propagators with momentum $ l- \Delta  $ and $ p_A - l $ in the right of Figure \ref{fig:problematic-diagram}. They pinch the $ l^{-} $ component in the DGLAP region in the nucleon sector (i.e.~when $ p_{A}^{+}>0 $), since (note that $  \Delta ^{+}< 0 $ in the kinematics we use)
\begin{align}
	 \left( l- \Delta  \right) ^{2} + i\epsilon =0 \Rightarrow l^{-} =  {\cal O } ( \lambda ^2) - i \epsilon\,, \qquad
	 \left( p_A - l  \right) ^{2} + i\epsilon =0 \Rightarrow l^{-} =  {\cal O } ( \lambda ^2) +  i \epsilon \,.
\end{align}
We now consider the propagators with momenta $ k $ and $ p_{ \pi }-k $. They pinch the  $ k^{+} $ component since
\begin{align}
	  k ^2 +i\epsilon =0 \Rightarrow k^{+} =  {\cal O } ( \lambda )- \mathrm{sgn}(k^{-}) i\epsilon \,,\qquad
	  \left(p_{ \pi }- k \right)^2 +i\epsilon =0 \Rightarrow k^{+} =  {\cal O } ( \lambda^{2} )+i\epsilon \,,
\end{align}
with $ k^{-} >0 $ (ERBL region in the pion sector). Now, we consider the propagators with momenta $ k+l $ and $ q-k-l $. They pinch the $ l^{+} $ component since
\begin{align}
	 \left( k+l \right) ^2 + i\epsilon  \Rightarrow l^{+} =  {\cal O } ( \lambda ) -  \mathrm{sgn}(k^{-}) i\epsilon \,,\qquad
	 \left( q-k-l \right) ^2 + i\epsilon  \Rightarrow l^{+} =  {\cal O } ( \lambda ) + i\epsilon \,,
\end{align}
as well as the $ k^{-} $ component,
\begin{align}
 \left( k+l \right) ^2 + i\epsilon  \Rightarrow k^{-} =  {\cal O } ( \lambda ) - \mathrm{sgn}(l^{+}) i\epsilon \,,\qquad
\left( q-k-l \right) ^2 + i\epsilon  \Rightarrow k^{-} =  {\cal O } ( \lambda ) + i\epsilon \,,
\end{align}
with $ l^{+}>0 $ (DGLAP region). Hence, we find that the components of loop momenta $ l^{+},\, l^{-},\,k^{+},\,k^{-} $ are all pinched when the gluon loop momentum $ l $ has a Glauber scaling, and the quark loop momentum $ k $ has a soft scaling. A peculiar property of this Glauber pinch is that it is the result of a conspiracy between \textit{two} loop momentum integrations, which is why it is important that all of the \textit{four} previously mentioned momentum components have to be pinched \textit{simultaneously}.

In \cite{Nabeebaccus:2023rzr}, we further show that a power counting analysis of this Glauber pinch, shows that the amplitude scales as $  \lambda ^{1} $, which is the same as the collinear pinch. Finally, we refer the interested reader to our paper \cite{Nabeebaccus:2023rzr} for a detailed explanation as to why the Ward identities do not lead to a suppression of the amplitude for the Glauber pinch wrt that corresponding to the collinear pinch.

The Glauber pinch that is observed in the photoproduction of a $  \pi ^{0}\gamma  $ pair is similar to the case of a \textit{double diffractive process}, e.g.~ $ pp \to pp\gamma  \pi ^{0} $, where the Glauber gluon is pinched between two collinear sectors defined by the two pairs of incoming and outgoing nucleons. Instead, here, the Glauber gluon corresponds to one of the active partons, and is pinched between one pair of collinear nucleons, and a \textit{soft line} joining the outgoing pion and the incoming photon.

Such a Glauber pinch also exists for the gluon exchange channel for the crossed process of $  \pi ^{0}N \to \gamma \gamma N $ which is discussed in \cite{Qiu:2022bpq}. Still, we emphasise that the channels we have considered in Section \ref{sec:photon-meson-photoproduction}, namely the photoproduction of a $  \pi ^{\pm}\gamma  $ pair and of a $  \rho \gamma  $ pair, are completely safe from the factorisation breaking discussed here. This is due to the fact that the two gluon exchange channel is forbidden, either on the basis of electric charge conservation or charge conjugation symmetry.

\section{Conclusions}

Our results show that the exclusive photoproduction of a photon-meson pair is a very promising channel to study quark GPDs. In addition to providing an extra channel to probe quark GPDs, it is one of the very few processes that can be used to extract chiral-odd GPDs at the leading twist, by choosing the outgoing meson to be a transversely polarised $  \rho  $-meson. We estimated the statistics at various experiments using our calculated cross sections, and we found that it is feasible to measure the process at JLab, COMPASS, future EIC and LHC in ultraperipheral collisions. In particular, our estimates for JLab, including an analysis of the detection of the outgoing photon, shows that the expected number of events is of the order of $ 10^5 $.

We further showed that for corresponding processes which have contributions from the gluon GPD channel, collinear factorisation breaks down. This is due to the presence of Glauber gluon pinches, which occurs when one of the active gluons that joins the nucleon sector to the hard part itself becomes a Glauber gluon. This point is evidenced by the fact that an explicit calculation of the gluon GPD channel to $  \pi ^{0}\gamma  $ pair photoproduction, assuming collinear factorisation, leads to a \textit{divergent} amplitude, already at the leading order. We emphasize that such a problem with collinear factorisation exist \textit{only} for channels where the 2-gluon exchange in the $ t $-channel is allowed.

\section*{Acknowledgements}

This work was supported by the GLUODYNAMICS project funded by the ``P2IO LabEx (ANR-10-LABEX-0038)'' in the framework ``Investissements d’Avenir'' (ANR-11-IDEX-0003-01) managed by the Agence Nationale de la Recherche (ANR), France. This work was also supported in part by the European Union’s Horizon 2020 research and innovation program under Grant Agreements No. 824093 (Strong2020). This project has also received funding from the French Agence Nationale de la Recherche (ANR) via the grant ANR-20-CE31-0015 (``PrecisOnium'')  and was also partly supported by the French CNRS via the COPIN-IN2P3 bilateral agreement. The work of L.S. is supported by the grant 2019/33/B/ST2/02588 of the National Science Center in Poland. L.S. thanks the P2I - Graduate School of Physics of Paris-Saclay University for support. J.S. was supported in part by the Research Unit FOR2926 under grant 409651613, by the U.S. Department of Energy through Contract No. DE-SC0012704 and by Laboratory Directed Research and Development (LDRD) funds from Brookhaven Science Associates. This work is also supported by the Croatian Science Foundation project IP-2019-04-9709.

\bibliographystyle{utphys}
\bibliography{../../../masterrefs.bib}



\end{document}